\documentclass[12pt,a4]{article} 
\begin{document} 
\title{\bf Accelerated Universe from   
Modified Chaplygin Gas and Tachyonic Fluid   
} 
\author{ H. B. Benaoum \footnote{hbenaoum@physics.syr.edu}
  \\
Physics Department, Syracuse University, \\
Syracuse, NY~13244--1130,~ USA }
\date{}
\maketitle
~\\
\abstract{ A cosmological model with an exotic fluid is investigated. 
We show that the equation of state of this ``modified Chaplygin'' gas 
can describe the current accelerated expansion of the universe. 
We then reexpress it as FRW cosmological model containing a scalar 
field $\phi$ and find its self--interacting potential. 
Moreover motivated by recent works of Sen~\cite{sen1, sen2} and 
Padmanbhan~\cite{pad} on tachyon field theory, a map for this exotic 
fluid as a normal scalar field $\phi$ with Lagrangian 
${\cal L}_{\phi} = \frac{\dot{\phi}^2}{2} - U (\phi)$ to the 
tachyonic field $T$ with Lagrangian ${\cal L}_{T} = - V(T) 
\sqrt{1 - \dot{T}^2 }$ is obtained. 
} 
~\\ 
~\\
~\\
~\\
\newpage
~\\  
\section{Introduction:}  

Recent measurements of redshift and luminosity-distance relations of 
type $Ia$ supernovae indicate that the expansion of
the universe is accelerating~\cite{per, rie}. This appears to be in 
strong disagreement with the standard picture of a matter dominated 
universe.  
 
These observations can be accommodated theoretically by postulating  
that certain exotic matter with negative pressure dominates  
the present epoch of our universe. Such exotic matter has been called 
{\em Quintessence} and behaves like a vacuum field energy with repulsive 
(anti--gravitational) character arising from the negative pressure. 

Negative pressure leading to an accelerating universe can also 
be obtained in a Chaplygin gas cosmology~\cite{pas},
in which the matter is taken to be a perfect fluid 
obeying an exotic equation of state. This cosmological model has some 
interesting properties. In particular, the Chaplygin gas behaves as 
pressureless fluid for small values of the scale factor and as a 
cosmological constant for large values of the scale factor which tends to 
accelerate the expansion.  
 
Another interesting feature of the Chaplygin gas equation of state 
is its connection to string theory via a brane interpretation.  
It was shown in~\cite{nao} that this kind of exotic equation of state can 
be obtained from the Nambu-Goto action for a $D$-brane moving in a $(D+2)$-
dimensional spacetime in light-cone parametrization. 
 
In this paper we consider a class of equations of state  
that interpolate between standard fluids at high energy densities
and Chaplygin gas fluids at low energy densities. We call 
this type of equation of state the ``modified'' Chaplygin gas. In the 
next section, we study the behavior of this modified Chaplygin gas and show 
that at large cosmological scales it could account for the current 
observations of the acceleration of the universe. In section 3, we 
investigate the possibility of using Sen's~\cite{sen1, sen2,sen3} idea of 
a rolling tachyon to describe the current acceleration of the present 
epoch where the tachyon fluid is considered as a candidate for 
the modified Chaplygin gas equation of state. 

\section{Modified Chaplygin gas:}

Within the framework of FRW, we study a model based on ( modified ) 
Chaplygin gas where our principal assumption is that the energy density 
$\rho$ and pressure $p$ are related by the following equation of 
state :  
\begin{eqnarray}
p & = & A \rho - \frac{B}{\rho^n} ,~~~~~~n \geq 1  
\end{eqnarray}
where  $A$ and $B$ positive constants. \\  
We see that when $B = 0$, it reduces to the standard equation of state 
of perfect fluid, 
\begin{eqnarray}
p & = & A~ \rho 
\end{eqnarray}
whereas when $A = 0$, it corresponds to an exotic background fluid, 
Chaplygin gas, described by an equation of state : 
\begin{eqnarray}
p & = & -~ \frac{B}{\rho}~,~~~~\mbox{ for n = 1} 
\end{eqnarray}
In (1), the two terms start to be of the same order when the pressure 
vanishes ( i.e. $p = 0$ ). In this case, the fluid has pressureless 
density $\rho_0$, corresponding to some cosmological scale $a_0$ , 
\begin{eqnarray}
\rho_0 & = & \rho^{n+1} ( a_0 )~=~ \frac{B}{A} 
\end{eqnarray}
Now we consider a $D$-dimensional FRW spacetime with scale factor $a(t)$ 
and metric :
\begin{eqnarray}
d s^2 & = & - d t^2 + a{^2}(t)~ d \Omega_k{^2}
\end{eqnarray} 
where $d \Omega_k{^2}$ is the metric of the maximally symmetric $( D - 1)$--
space with curvature $k$, for $k = -1,0, +1$ . \\  
If this space time is filled with a fluid of energy density $\rho$ and 
pressure $p$, then conservation of energy momentum tensor 
$\nabla^{\mu} T_{\mu \nu} = 0$, gives, 
\begin{eqnarray} 
\dot \rho + (D-1) H ( \rho + p ) & = & 0 
\end{eqnarray}   
The Friedmann equation for the scale factor $a$ is, 
\begin{eqnarray}
H^2 & = & \left( \frac{\dot{a}}{a} \right)^2~ =~ 
\frac{2~\rho}{(D-1) (D-2)} - \frac{k}{a^2} 
\end{eqnarray}
These two equations imply that :
\begin{eqnarray}
\frac{\ddot{a}}{a} & = & \dot{H} + H^2~ =~ -~
\frac{(D-1)~p + (D-3)~\rho}{(D-1)(D-2)} 
\end{eqnarray}
A solution to this equation is obtained as follows. We define 
$W = a^{(D - 1) (A + 1)}$ and a rescaled density 
$\bar{\rho} = \rho W$, then (6) becomes, 
\begin{eqnarray}
\dot{\bar{\rho}} - \frac{B}{A + 1}~\frac{W^n \dot{W}}{\bar{\rho}^n} & = & 0 
\end{eqnarray} 
The latter has a solution of the form :  
\begin{eqnarray}
\frac{\bar{\rho}^{n+1}}{n + 1} & = & \frac{B}{A+1} \frac{W^{n+1}}{n+1}~ 
+~ \frac{C}{n+1}  
\end{eqnarray}
where $C$ is a constant of integration. \\
The density will be : 
\begin{eqnarray}
\rho & = & \left( \frac{B}{A+1} 
+ \frac{C}{W^{n+1}} \right)^{\frac{1}{n+1}} 
\end{eqnarray} 
The constant of integration $C$ can be expressed in terms of the 
cosmological scale $a_0$,~( i.e. $W_0 = a_0^{(D -1 )(A+1)}$~) 
where the fluid has vanishing pressure :  
\begin{eqnarray} 
C & = & \frac{B}{A+1}~ \frac{W_0^{n+1}}{A} 
\end{eqnarray} 
The energy density $\rho$ will be : 
\begin{eqnarray}
\rho & = & \left( \frac{B}{A+1} \right)^{\frac{1}{n+1}}~ \left( 1 + 
\frac{1}{A}~ \left( \frac{W_0}{W} \right)^{n+1} \right)^{\frac{1}{n+1}}  
\end{eqnarray} 
We see that for a large values of the scale factor $a$, i.e. 
$W >> W_0$, we get : 
\begin{eqnarray} 
\rho & \simeq & \left( {\frac{B}{A+1}} \right)^{\frac{1}{n+1}}  \nonumber \\
p & \simeq & - \left({\frac{B}{A+1}} \right)^{\frac{1}{n+1}} = - \rho 
\end{eqnarray} 
which correspond to an empty universe with a cosmological constant 
$\left( {\frac{B}{A+1}} \right)^{\frac{1}{n+1}}$ (i.e. a de Sitter space ). \\
On the other hand, for small scale factor $a$, 
i.e.  $W << W_0$, we get :  
\begin{eqnarray} 
\rho & \simeq & \frac{\root n+1 \of C}{W} \nonumber \\
p & \simeq & A \frac{\root n+1 \of C}{W} = A \rho 
\end{eqnarray} 
which correspond to universe dominated by an equation of state 
$p = A \rho$. This shows that this model interpolates between a universe 
dominated by matter phase with equation of state $p = A \rho$ and 
a de Sitter phase $p \simeq - \rho$. \\ 
Moreover, expanding equation (11) and (1) to the subleading terms at large 
cosmological constant, we obtain the following expressions for the energy 
and the pressure : 
\begin{eqnarray}
\rho & \simeq & \left( {\frac{B}{A+1}} \right)^{\frac{1}{n+1}}~
\left( 1 + \frac{1}{ ( n+1) A} 
\left( \frac{W_0}{W} \right)^{n+1} \right) \nonumber \\
p & \simeq & \left( {\frac{B}{A+1}} \right)^{\frac{1}{n+1}}~ 
\left( - 1 + \frac{A + n (A + 1)}{( n+1) A} 
\left( \frac{W_0}{W} \right)^{n+1} \right)  
\end{eqnarray} 
These correspond to the mixture of a cosmological constant 
$\left( \frac{B}{A+1} \right)^{\frac{1}{n+1}}$ and a type of matter 
described by an equation of state : 
\begin{eqnarray}
p & = & \left( n + A (n + 1) \right)~ \rho 
\end{eqnarray}  
The equation of state parameter takes the form :
\begin{eqnarray}
\omega & = & \frac{p}{\rho}~=~ -~ 
\frac{1 - \left( \frac{W_0}{W} \right)^{n+1}}{1 + \frac{1}{A} 
\left( \frac{W_0}{W} \right)^{n+1}} 
\end{eqnarray}
which ranges over $-1 < \omega < A$ , depending on the cosmological scale 
$a$, 
\begin{eqnarray}
\omega & = & - 1 \mbox{ for}~~~W \gg W_0   \nonumber \\
\omega & = & 0  \mbox{ for}~~~W = W_0 \nonumber \\
\omega & = & A \mbox{ for}~~~W \ll W_0 
\end{eqnarray}
The speed of sound $c_s$ is defined as, 
\begin{eqnarray} 
c_s^2 & = & \frac{\delta p}{\delta \rho}~=~ \frac{\dot{p}}{\dot{\rho}} 
\end{eqnarray} 
Now by computing $\dot{\omega}$, we obtain : 
\begin{eqnarray} 
\dot{\omega} & = & \left( c_s^2 - \omega \right)~ \frac{\dot{\rho}}{\rho}  
\end{eqnarray} 
It gives the following expression for the speed of sound :   
\begin{eqnarray} 
c_s^2 & = & \omega + \rho \frac{d \omega}{d \rho} ~ = ~ 
\frac{n + A (n +1) + \left( \frac{W_0}{W} \right)^{n+1}}{1 + \frac{1}{A} 
\left( \frac{W_0}{W} \right)^{n+1}} 
\end{eqnarray} 
which implies that $c_s^2$ is always positive and hence there is no concern 
about imaginary speed of sound. \\ 
Moreover it has the following asymptotic limit, 
\begin{eqnarray}
c_s^2 & = & n + A ( n + 1)~~~~~\mbox{for}~~W \gg W_0  \nonumber \\
c_s^2 & = & A ( n +1 )~~~~~~\mbox{for}~~W = W_0  \nonumber \\
c_s^2 & = & A~~~~~~~~\mbox{for}~~W \ll W_0   
\end{eqnarray}  
The speed of sound never exceeds that of light for smaller scale $a$ 
(i.e. $W$ ) or of the order of the scale $a_0$ where the pressure 
vanishes, provided that $A < \frac{1}{n + 1}$ and will exceed it 
for large scale compared to $a_0$. \\  
~\\
A way to describe this cosmological model from field theoretical 
point of view is to introduce a scalar field $\phi$ and self-interacting 
potential $U(\phi)$ with the following Lagrangian~\cite{bar} : 
\begin{eqnarray}
{\cal L}_{\phi} & = & \frac{\dot{\phi}^2}{2} - U(\phi) . 
\end{eqnarray}
The energy momentum tensor contributed by the scalar field $\phi$ is 
identical to a fluid with energy density $\rho_{\phi}$ and pressure 
$p_{\phi}$ given by : 
\begin{eqnarray}
\rho_{\phi} & = & \frac{\dot{\phi}^2}{2} + U(\phi) =~ \rho  \nonumber \\
p_{\phi} & = & \frac{\dot{\phi}^2}{2} - U(\phi) =~ A \rho - \frac{B}{\rho^n} .
\end{eqnarray} 
It follows that : 
\begin{eqnarray}
\dot{\phi}^2 & = & ( 1 + \omega_{\phi} )~ \rho_{\phi}  \nonumber \\
U(\phi) & = & \frac{1}{2}~( 1 - \omega_{\phi} )~\rho_{\phi} 
\end{eqnarray}
Now since $\dot{\phi} = \phi' \dot{W}$ where the prime denotes derivation 
with respect to $W$ and $\dot{W} = (D -1) (A + 1) H W$, we get : 
\begin{eqnarray}
\phi'{^2} & = & \frac{D -2}{2 (D -1) (A+1)^2}~\frac{1 + \omega}{W^2} 
\end{eqnarray} 
Here we have used (7) for the Hubble constant and guided by the cosmic 
microwave background CMB data which is strongly consistent with a flat 
universe, we have restricted ourselves to the flat case $k = 0$. \\ 
A use of (18) for $n=1$, gives : 
\begin{eqnarray}
\phi' & = & \sqrt{\frac{D -2}{2 (D -1) A (A +1)}}~ 
\frac{\frac{W_0}{W^2}}{\sqrt{1 + \frac{1}{A} 
\left( \frac{W_0}{W} \right)^2}} \nonumber \\
U \left( \phi \right) & = & \frac{1}{2} 
\left(\frac{B}{A+1} \right)^{\frac{1}{2}}~ \frac{2 + \frac{1 -A}{A} 
\left( \frac{W_0}{W} \right)^2}{\sqrt{1 + \frac{1}{A} \left( 
\frac{W_0}{W} \right)^2}} 
\end{eqnarray}  
The first equation can be integrated easily which gives : 
\begin{eqnarray}
A \left( \frac{W}{W_0} \right)^2 & = & 
\frac{4 e^{2 \alpha \phi}}{\left( 1 - e^{2 \alpha \phi} \right)^2}~ = ~ 
\frac{1}{ sh^2 (\alpha \phi)} 
\end{eqnarray}
where $\alpha = \sqrt{\frac{2 (A +1) (D -1)}{D -2 }}$. \\
We note that for larger scales (i.e. $W \gg W_0$ ), 
the scalar vanishes (i.e. $\phi = 0$ ) 
and becomes infinite (i.e. $\phi \rightarrow \ + 
\infty$ ) for small scales 
( $W \ll W_0$ ). \\
Finally, by substituting the latter expression in our previous equations, 
we can write all our physical quantities $\rho_{\phi}, p_{\phi}$ and 
$\omega_{\phi}$ in terms of the scalar field $\phi$ as, 
\begin{eqnarray}
\rho_{\phi} & = & \left( \frac{B}{A + 1} \right)^{\frac{1}{2}}~
\cosh (\alpha \phi) \nonumber \\
p_{\phi} & = & \left( \frac{B}{A + 1} \right)^{\frac{1}{2}}~
\left( A~\cosh (\alpha \phi) - 
\frac{A+1}{\cosh (\alpha \phi)} \right) \nonumber \\
\omega_{\phi} & = & -~\frac{1 - A~\sinh^2 (\alpha \phi )}{\cosh^2 (\alpha \phi)}
\end{eqnarray}
Notice that these physical quantities do not depend on the intermediate 
constant $W_0$ (i.e. constant of integration $C$ ). \\
Finally, we get the following potential which has a simple form :
\begin{eqnarray}
U ( \phi) & = & \frac{1}{2} \left( \frac{B}{A + 1} \right)^{1/2}~ 
\left( \frac{1 + A}{\cosh (\alpha \phi)} + (1  - A) \cosh (\alpha \phi) \right) 
\end{eqnarray} 
\section{Rolling Tachyon :}  
Recently, some works have appeared in the literature which 
study the tachyon field cosmology~\cite{gib1}--\cite{shiu}. 
These follow from Sen's~\cite{sen1, sen2} idea   
which suggests that the tachyon condensate of string theory in a 
gravitational background may be described by an effective field theory 
with an action of the form : 
\begin{eqnarray} 
S & = & \int d^D x \sqrt{- g}~\left( R - V(T)~\sqrt{ 1 + g^{\mu \nu} 
\nabla_{\mu} T \nabla_{\nu} T} ~  \right) 
\end{eqnarray} 
where $T$ is the tachyon field, $V(T)$ is the tachyon potential and 
$g_{\mu \nu}$ is the FRW metric spacetime. The tachyon potential $V(T)$ 
has two extremal points. The extremal point $T=0$ is a maximum with 
$V(T = 0) = V_0$ as the tension of some unstable bosonic $D$-brane and 
the extremal point $T \rightarrow T_0$ is a minimum where the 
potential vanishes. \\ 
The energy momentum tensor is :
\begin{eqnarray}
T_{\mu \nu} & = & - \frac{2 \delta S}{\sqrt{- g} \delta g^{\mu \nu}}~ = ~ 
- V(T) \sqrt{ 1 + g^{\mu \nu} \nabla_{\mu}T \nabla_{\nu} T }~ g_{\mu \nu} 
+ V(T) \frac{\nabla_{\mu} T \nabla_{\nu} T}{
\sqrt{ 1 + g^{\mu \nu} \nabla_{\mu}T \nabla_{\nu} T }}  \nonumber \\
& = & p_T~ g_{\mu \nu} + \left( p_T + \rho_T \right)~ u_{\mu} u_{\nu} 
\end{eqnarray}
where the 4-velocity $u_{\mu}$ is : 
\begin{eqnarray}
u_{\mu} & = & \frac{- \nabla_{\mu} T}{\sqrt{ - g^{\mu \nu} \nabla_{\mu} T   
\nabla_{\nu} T }}
\end{eqnarray}  
with $u^{\mu} u_{\mu} = - 1$ . \\
Then the energy density $\rho_T$ and the pressure $p_T$ of the tachyon field 
which we consider homogeneous but time dependent is ( i.e. $\nabla_i T = 0$ ), 
\begin{eqnarray} 
\rho_T & = & \frac{V(T)}{\sqrt{1 - \dot{T}^2}} \nonumber \\
p_T & = & - V(T) \sqrt{ 1 - \dot{T}^2 } 
\end{eqnarray}
The condition of accelerating universe (i.e. $\frac{\ddot{a}}{a} > 0$ ) 
requires that : 
\begin{eqnarray}
\dot{T}^2 < \frac{2}{D - 1 } 
\end{eqnarray}
Then the equation of state parameter is : 
\begin{eqnarray}
\omega_T & = & \frac{p_T}{\rho_T}~=~ - \left(1 -  \dot{T}^2 \right) 
\end{eqnarray} 
Now consider the pressure $p_{\phi} = \frac{\dot{\phi}^2}{2} - U (\phi)$ and 
the energy density $\rho_{\phi} = \frac{\dot{\phi}^2}{2} + U (\phi)$ for 
the scalar field that describes the modified Chaplygin gas for 
the case $n=1$ and rewrite them as : 
\begin{eqnarray}
p_{\phi} & = & 
- U (\phi)~ \left( 1 - \frac{\dot{\phi}^2}{2~U (\phi)} \right) \nonumber \\ 
\rho_{\phi} & = & U (\phi)~\left( 1 + \frac{\dot{\phi}^2}{2~U (\phi)} \right) 
\end{eqnarray} 
In the approximation $\frac{\dot{\phi}^2}{U (\phi)} \ll 1$, the above 
pressure and energy density can expressed in the following form : 
\begin{eqnarray}
p_{\phi} & \simeq & - U (\phi)~ 
\sqrt{1 - \frac{\dot{\phi}^2}{U (\phi)}} \nonumber \\ 
\rho_{\phi} & \simeq & 
\frac{U (\phi)}{\sqrt{1 - \frac{\dot{\phi}^2}{U (\phi)}}} 
\end{eqnarray}
Mapping the pressure $p_{\phi}$ and energy density $\rho_{\phi}$ for 
the scalar field in this approximation with the corresponding tachyon 
field $p_T$ and $\rho_T$ (35), we obtain that : 
\begin{eqnarray}
\dot{T}^2 & = & \frac{\dot{\phi}^2}{U (\phi)}~ = ~ 
2~ \left( \frac{p_{\phi} + 
\rho_{\phi}}{\rho_{\phi} - p_{\phi}}  \right) \nonumber \\
V (T) & = & U (\phi)~ = ~ \frac{1}{2} \left( \rho_{\phi} - p_{\phi} \right) 
\end{eqnarray} 
It follows then that : 
\begin{eqnarray}
\dot{T} & = & \sqrt{2~ \left( 
\frac{p_{\phi} + \rho_{\phi}}{\rho_{\phi} - p_{\phi}} \right) }~=~ 
\sqrt{2~ \left(\frac{1 + \omega_{\phi}}{1 - \omega_{\phi}} \right) }
\end{eqnarray} 
We see that the above approximation $\frac{\dot{\phi}^2}{U( \phi)} \ll 1$ 
is valid only when $\omega_{\phi} \ll \frac{- 1}{3}$. It implies that the 
fluid with pressure $p_{\phi}$ and energy density $\rho_{\phi}$ has an 
equation of state $p_{\phi} \ll - \frac{1}{3} \rho_{\phi}$ which can 
describe the current observations of accelerating universe. Moreover for 
large values of the scale factor $W \gg W_0$ ( i.e. $\phi \rightarrow 0$ ), 
$\dot{T} = 0$ which corresponds to $\omega_T = -1$. \\ 
Equation (41) can be rewritten as : 
\begin{eqnarray}
\dot{T} & = & - \sqrt{\frac{D-2}{( D-1) (1 - \omega_{\phi}^2 )}}~ 
\frac{\dot{\rho_{\phi}}}{\rho_{\phi}^{3/2}}  
\end{eqnarray} 
where we have used equation (6) and (7). \\ 
Then the tachyon field $T$ can be found in terms of the energy density 
$\rho_{\phi}$ for the scalar field as : 
\begin{eqnarray}
T & = & - \sqrt{\frac{D-2}{D-1}}~ \int 
\frac{d \rho_{\phi}}{\sqrt{1 - \omega_{\phi}^2 }~ \rho_{\phi}^{3/2}}  
\end{eqnarray} 
An approximate solution to this integral can be found by replacing 
$\omega_{\phi}$ with its mean value $< \omega_{\phi} >$, 
\begin{eqnarray}
T & \simeq & - \sqrt{\frac{D-2}{(D-1)~(1 - < \omega_{\phi} >^2 )}}~ \int 
\frac{d \rho_{\phi}}{\rho_{\phi}^{3/2}}  \nonumber \\
& = & 2~ \sqrt{\frac{D-2}{(D-1)~(1 - < \omega_{\phi} >^2 )}}~ 
\frac{1}{\sqrt{\rho_{\phi}}} 
\end{eqnarray}
This means that : 
\begin{eqnarray}
\rho_{\phi} & \simeq & 4~ 
\left( \frac{D-2}{(D-1)~(1 - < \omega_{\phi} >^2 )} \right)~ \frac{1}{T^2} 
\end{eqnarray} 
and the potential $V(T)$ for the tachyon field can be expressed as, 
\begin{eqnarray}
V(T) & = & \frac{1}{2} \left( \rho_{\phi} - p_{\phi} \right)~ = ~ 
\frac{1}{2} \left(~ ( 1 - A )~ \rho_{\phi} 
+ \frac{B}{\rho_{\phi}} \right)  \nonumber \\
& \simeq & \frac{1}{2} 
\left(~ \frac{ 4 (1 - A) (D -2)}{(D-1) (1 - < \omega_{\phi} >^2 )}~
\frac{1}{T^2} +   
\frac{B (D - 1) (1 - < \omega_{\phi} >^2 )}{4 (D-2) }~ T^2 \right) 
\end{eqnarray}  
Even though the tachyonic potential $V(T)$ has different form than 
the scalar potential $U( \phi )$. They lead to the same cosmological 
evolution that is~ :~ an accelerated expansion for the universe. \\
This was made possible because of the mapping that exists between normal 
scalar field $\phi$ with potential $U( \phi)$ and tachyonic field $T$ 
with potential $V(T)$.  \\

Before we close, we remark that the mapping between scalar field $\phi$ and 
tachyon field $T$ which is valid for small gradient 
$\dot{T}^2 = \frac{\dot{\phi}^2}{U (\phi)} \ll 1$, suggests that the 
tachyonic Lagrangian ( which is the pressure in this case ), 
\begin{eqnarray}
p_T & = & - V(T)~ \sqrt{1 - \dot{T}^2 } \nonumber \\
& \simeq & - V(T)~ ( 1 - \frac{\dot{T}^2}{2} ) 
\end{eqnarray}
is identical to the $k$--essence Lagrangian~\cite{stein}, 
$p_k (T, \frac{\dot{T}^2}{2} ) = - V(T)~ \tilde{p} ( \frac{\dot{T}^2}{2})$. \\ 
This implies that the mapping between $\phi$ and $T$ can be reinterpreted now 
as a mapping between normal scalar field $\phi$ and $k$--essence field. 
\section{Conclusions : }
We have considered an exotic fluid and studied its cosmological 
implications. Such an exotic fluid obeys an equation of state,
we called it '' modified Chaplygin '' equation of state, which has some 
interesting properties. \\ 
It has been shown here that this '' modified Chaplygin '' 
gas of state describes the evolution of a universe from a phase dominated 
by an equation of state $p = A \rho$ for small values of the scale 
cosmological factor to a phase dominated by a cosmological 
constant $\left( \frac{B}{A+1} \right)^{\frac{1}{n+1}}$ for large values 
of the scale factor. \\
We have then described this '' modified Chaplygin '' gas as FRW 
cosmological model having a scalar field and found its self-interacting 
potential. \\ 
Furthermore we have discussed the cosmological evolution of the tachyon field 
in gravitational background to drive the accelerated expansion of the 
universe and obtained a mapping between a normal scalar field with its 
potential and a tachyonic field with its corresponding potential.  
\section*{Acknowledgment }
I would like to thank L. Kofman, S. Nasri and G. Shiu for
 reading the manuscript and 
sending me their comments. 


\begin{thebibliography}{99}  
\bibitem{per} 
S. Perlmutter et al., Bull. Am. Astron. Soc. {\bf 29}, 1351 (1997) ; 
Astrophys. J. {\bf517}, 565 (1999) .
\bibitem{rie}
A.G. Riess et al., Astron. J. {\bf 116}, 1009 (1998) ; 
P. Garnavich et al., Astrophys. J. {\bf 493}, L53 (1998) ; 
B.P. Schmidt et al. Astrophys. J. {\bf 507}, 46 (1998) . 
\bibitem{pas}
A. Kamenshchik, U. Moschella and V. Pasquier, Phys. Lett. {\bf B511}, 265 
(2001) . 
\bibitem{nao}
N. Ogawa, Phys. Rev. {\bf D62}, 085023, (2000) ; A. Kamenshchik et al., 
Phys. Lett. {\bf B487}, 7 (2000) ; N. Bilic and al., astro-ph/0111325 ; 
M.C. Bento et al., gr-qc/0202064 ; J.C. Fabris et al., astro-ph/0203441 .
\bibitem{sen1} 
A. Sen, '' Rolling Tachyon '', hep-th/0203211 . 
\bibitem{sen2} 
A. Sen, '' Tachyon Matter'', hep-th/0203256 .   
\bibitem{sen3} 
A. Sen, '' Field Theory of Tachyon Matter '', hep-th/0204143 .
\bibitem{bar} 
J.D. Barrow, Phys. Lett. {\bf B235}, 40 (1990) ; 
J.D. Barrow and P. Saich, Class. Quantum Grav. {\bf 10}, 279 (1993) .
\bibitem{gib1}
A.W. Gibbons, '' Cosmological Evolution of the Rolling Tachyon '', 
hep-th/0204008 . 
\bibitem{fai}
M. Fairbian and M.H.G. Tytgat, '' Inflation from a Tachyon Fluid ? '', 
hep-th/0204070 .   
\bibitem{muko}
S. Mukhoyama, '' Brane cosmological driven by the rolling tachyon '', 
hep-th/0204084 .
\bibitem{fei}
A. Feinstein, '' Power-Law Inflation from the Rolling Tachyon '', 
hep-th/0204140 .
\bibitem{pad} 
T. Padmanabhan, '' Accelerated expansion of the universe driven by 
tachyonic matter '', hep-th/0204150 . 
\bibitem{frol}
A. Frolov, L. Kofman and A.A. Starobinsky, '' Prospects and problems 
of tachyon matter cosmology '', hep-th/0204187 . 
\bibitem{deba} 
D. Choudhury et al., '' On the cosmological relevance of the tachyon '', 
hep-th/0204204 . 
\bibitem{zhou} 
Xin-zhou Li, Jian-gang Hao and Dao-jun Liu, '' Can quintessence be the 
rolling tachyon ? '', hep-th/0204252 . 
\bibitem{shiu}
G. Shiu and I. Wasserman, '' Cosmological constraints on tachyon matter '', 
hep-th/0205003 . 
\bibitem{stein}
C. Armendariz--Picon, V. Mukhanov and P.J. Steinhardt, Phys. Rev. D {\bf 63} 
(2001) 103510; C. Armendariz--Picon, T. Damour and V. Mukhanov, Phys. Lett. 
{\bf B458}, (1999) 209; T. Chiba, T. Okabe and M. Yamaguchi, Phys. Rev. 
{\bf D62}, (2000) 023511 . 
\end{thebibliography}
\end{document}